\documentclass[12pt,twoside]{article}
\usepackage{amssymb}
\usepackage{amsmath,amsthm}
\usepackage{latexsym}
\usepackage{graphicx}
\usepackage{psfrag}
\usepackage{subfig}
\begin{document}
\begin{center}
{\Large {\bf Notes on coherent backscattering from a random
potential\bigskip\bigskip\\}}
{\large{Herbert Spohn}}\footnote{{\tt spohn@ma.tum.de}}\medskip\\
Zentrum Mathematik and Physik Department, TU M\"{u}nchen,\\
D - 85747 Garching, Boltzmannstr. 3, Germany
\end{center}\bigskip\bigskip\bigskip\bigskip
{\bf Abstract}. We consider the quantum scattering from a random
potential of strength $\lambda^{1/2}$ and with a support on the
scale of the mean free path, which is of order $\lambda^{-1}$. On
the basis of maximally crossed diagrams we provide a concise formula
for the backscattering rate in terms of the Green's function for the
kinetic Boltzmann equation. We briefly discuss the extension to wave
scattering.
\newpage
\section{Introduction}\label{sec.1}

The motion of a single quantum particle in a weak random potential
of strength $\lambda^{1/2}$ is well approximated by a kinetic
equation. More precisely, in the limit of small $\lambda$, the disorder averaged Wigner function $W$
is governed by the transport equation
\begin{eqnarray}\label{1.1}
&&\hspace{-26pt}\frac{\partial}{\partial t}W(q,p,t) )= -\nabla_p\,\omega(p)\cdot
\nabla_q W(q,p,t)\\
&&\hspace{10pt}+\rho(q)\int
dp'2\pi\delta(\omega(p)-\omega(p'))(2\pi)^{-3/2}
\widehat{\theta}(p-p')\big(W(q,p',t)-W(q,p,t)\big)\,.\nonumber
\end{eqnarray}
Here $\omega(p)$ is the kinetic energy (= dispersion relation),
$\omega(p)=p^2/2m$ for a nonrelativistic particle.
$\widehat{\theta}$ is the Fourier transform of the translation
invariant part of the covariance  of the random potential and
$\rho(q)^{1/2}$ is the spatially varying strength of the potential.
The mean free time and the mean free path are both of order
$\lambda^{-1}$ and Eq. (\ref{1.1}) is written on that scale. As a
side remark, the Wigner function integrated against a smooth test
function is self-averaging \cite{C05}. Thus (\ref{1.1}) holds even
for a typical realization of the random potential.

(\ref{1.1}) can be thought of as arising from the stochastic motion
of a fictitious classical particle: the particle moves along a
straight line according to its velocity and randomly changes its
momentum, respecting energy conservation, with a rate given through
the covariance of the potential. In this sense the wave nature of
the quantum motion is completely lost as $\lambda\to 0$. However, as
discovered by Langer and Neal \cite{LN06}, even for small $\lambda$
the wave character is still visible in the fine structure of the
average scattering rate. We imagine an incoming plane wave with wave
number $k$ scattering off the random potential and denote by
$\sigma_\mathrm{B}(k,k')$ the scattering rate from the incoming wave
number $k$ to the outgoing wave number $k'$ according to the
transport equation (\ref{1.1}) and by $\langle
\sigma_\lambda(k,k')\rangle_V$ the scattering rate for the
Schr\"{o}dinger equation averaged over the random potential. Then
\begin{equation}\label{1.2}
\lim_{\lambda\to 0}\lambda^2\langle
\sigma_\lambda(k,k')\rangle_V=\sigma_\mathrm{B}(k,k')\quad
\mathrm{for }\, k'\neq \pm k\,.
\end{equation}
Note that the total cross section of the scattering potential is of
order $\lambda^{-2}$. Therefore in (\ref{1.2})
one needs to balance
by the prefactor $\lambda^2$. For exact backscattering one finds that
\begin{equation}\label{1.3}
\lim_{\lambda\to 0}\lambda^2\langle
\sigma_\lambda(k,-k)\rangle_V=2\widetilde{\sigma}_\mathrm{B}(k,-k)\,,
\end{equation}
where the tilde indicates that from $\sigma_\mathrm{B}$ the paths
with only a single scattering event  have to be omitted. In fact,
the enhanced backscattering consists of a narrow peak of width
$\lambda$ centered at $k'=-k$. As to be discussed, one finds
\begin{equation}\label{1.4}
\lim_{\lambda\to 0}\lambda^2\langle
\sigma_\lambda(k,-k+\lambda\kappa)\rangle_V=\sigma_{\mathrm{back},k}(\kappa)\,.
\end{equation}

Properties (\ref{1.2}) and (\ref{1.4}) can be seen from an expansion
of the scattering amplitude in powers of the potential $V$. Summing
all ladder diagrams yields (\ref{1.2}), while (\ref{1.4}) results
from the maximally crossed diagrams. Of course, to actually prove
the limits (\ref{1.2}), (\ref{1.4}) one would have to establish that
the contribution of the very many remaining diagrams vanishes as
$\lambda\to 0$. This task will not be addressed in my notes. But I
point out to the reader that very complete and precise estimates
have been achieved in the recent work of Erd\"{o}s, Salmhofer, and
Yau \cite{ESY05,ESY06}. These notes are written, at least in part, with the aim to
encourage a similar kind of analysis for the scattering rate.

Coherent backscattering according to (\ref{1.4}) has been confirmed
experimentally including the shape of the peak \cite{WM85,AWM86}.
The experiment is done for light scattering, since light is so much
easier to manipulate than electrons. In (\ref{1.4}) averaging over
disorder is required. In a single scan no peak can be disentangled.
Thus one either repeats the experiment many times using samples of
disordered glass or, more elegantly, scatters from a turbid
solution. Then the Brownian motion of the suspended particles
provides the averaging for free. Excellent theoretical texts are
available \cite{BF94,S95,AW88}. They are striving for even more
refined information, as e.g. the variance of the conductance
fluctuations and the statistics of speckle patterns \cite{BF94}.
Does there remain then anything to be done on the level of
(\ref{1.4})?

In fact, I was looking for a concise expression of
$\sigma_{\mathrm{back},k}$ in terms of the transition probability
resulting from Eq.
(\ref{1.1}) and could not find it in the literature. My result is
given at the beginning of Section \ref{sec.5}. The required
computation I find sufficiently illuminating to be put into written
form.

In the literature (as far as I checked) the conventional approach is
to start from the average of the square of the Green's function and
then to extract from it the scattering. We proceed in a way which
looks more systematic to us. For every realization of the disordered
medium the scattering rate is given through stationary scattering
theory, at least in principle. The average scattering rate is then
expanded in the disorder strength.

The standard Born expansion for the scattering amplitude is covered
in Section \ref{sec.2} including the average over the random
potential. The scattering theory for (\ref{1.1}) is explained in
Section \ref{sec.3}. The most lengthy part of our contribution is
the summation of the ladder diagrams, while the maximally crossed
diagrams then easily follow, see Sections \ref{sec.4} and
\ref{sec.5}. We add the modifications required when the
Schr\"{o}dinger equation is replaced by a wave equation and close
with the diffusion approximation to (\ref{1.4}).

\section{Average scattering rate for electrons}\label{sec.2}
\setcounter{equation}{0}

Because it is slightly simpler and more familiar, we first discuss
the stationary scattering theory for an electron moving in a random
potential. The physically more relevant wave scattering will require
only minor modifications, see Section  \ref{sec.6}.

We consider the hamiltonian
\begin{equation}\label{2.1}
    H=-\frac{1}{2}\Delta+\lambda^{1/2}V(x)\,,\quad \lambda>0\,,
\end{equation}
acting on the Hilbert space $\mathcal{H}=L^2(\mathbb{R}^3,dx)$. In
Fourier space the Schr\"{o}dinger equation then reads
\begin{equation}\label{2.1a}
i\frac{\partial}{\partial
t}\widehat{\psi}(k,t)=\omega(k)\widehat{\psi}(k,t)+
(2\pi)^{-3/2}\lambda^{1/2}\int dk_1
\widehat{V}(k-k_1)\widehat{\psi}(k_1,t)\,,
\end{equation}
where we introduced the dispersion relation $\omega(k)$,
\begin{equation}\label{2.2}
\omega(k)=\frac{1}{2}k^2
\end{equation}
for a nonrelativistic particle of mass 1. The formalism is such that
general dispersion relations are allowed. $V(x)$ is a Gaussian
random potential with zero mean $\langle V(x)\rangle_V=0$,
$\langle\cdot\rangle_V$ denoting the average over $V$. The
statistics of $V(x)$ is translation invariant locally but $V(x)$
vanishes outside some bounded region. To achieve it we introduce the
smooth shape function $\rho:\mathbb{R}^3\to\mathbb{R}$, $\rho\geq
0$, such that the support of $\rho$, supp $\rho$, is a bounded set.
Then the covariance for $V(x)$ is given by
\begin{equation}\label{2.3}
\langle V(x)V(y)\rangle_V=\rho(\lambda
x)^{1/2}\theta(x-y)\rho(\lambda y)^{1/2}\,.
\end{equation}
$\theta(x)=\theta(-x)$ and $\theta$ is assumed to have a rapid
decay. Its Fourier transform satisfies $\widehat{\theta}(k)\geq 0$,
so to have a positive definite covariance.

In Fourier space the covariance (\ref{2.3}) is peaked near $k'=-k$
and we write
\begin{eqnarray}%XXX\label{2.4}
&&\hspace{-0pt}\langle \widehat{V}(k)\widehat{V}(-k+\lambda
k_1)\rangle_V \nonumber\\
&&\hspace{56pt}=(2\pi)^{-3/2}\int
dg\widehat{\theta}(g+k)\lambda^{-3}
\widehat{\rho^{1/2}}(\lambda^{-1}g)^\ast \lambda^{-3}
\widehat{\rho^{1/2}}(\lambda^{-1}g+k_1)\nonumber\\
&&\hspace{56pt}\cong\widehat{\theta}(k)\lambda^{-3}(2\pi)^{-3/2}\int
dg\widehat{\rho^{1/2}}(g)^\ast \widehat{\rho^{1/2}}(g+k_1)\nonumber\\
&&\hspace{56pt}=\widehat{\theta}(k)\lambda^{-3}(2\pi)^{-3/2}\int
dx\rho(x)e^{-ik_1\cdot
x}=\widehat{\theta}(k)\lambda^{-3}\widehat{\rho}(k_1)\,.
\end{eqnarray}
Therefore, in approximation, we set
\begin{equation}\label{2.5}
\langle \widehat{V}(k)\widehat{V}(k')\rangle_V=
\widehat{\theta}(k)\widehat{\rho}_\lambda(k+k')\,,
\end{equation}
where
$\widehat{\rho}_\lambda(k)=\lambda^{-3}\widehat{\rho}(\lambda^{-1}k)$.
Below we will always use (\ref{2.5}), with symmetry and positivity
restored through the $\mathcal{O}(\lambda)$ corrections.

We consider a weak potential, $\lambda\ll 1$. Then the mean free
path of the electron is $\mathcal{O}(\lambda^{-1})$. The shape
function $\rho$ enforces $V$ to vanish outside a bounded region on the scale
$\lambda^{-1}$. Thus the total cross section is of order
$\lambda^{-2}$.

For almost every realization of $V$ the hamiltonian $H$ has a
well-defined unitary scattering matrix $S_\lambda$, see \cite{RS79}
Section XI.6 for a discussion of stationary scattering theory. In
momentum space $S_\lambda$ has the kernel
\begin{equation}\label{2.6}
S_\lambda(k,k')=\delta(k-k')-2\pi i
T_\lambda(k,k')\delta(\omega(k)-\omega(k'))\,.
\end{equation}
Let us set $\varphi_k(x)=(2\pi)^{-3/2}e^{ik\cdot x}$. Then the
$T$-matrix is defined through
\begin{eqnarray}\label{2.4}
&&\hspace{-36pt}T_\lambda(k,k')=\lim_{\varepsilon\to 0}\langle
\varphi_k,\big(\lambda^{1/2}V-\lambda^{1/2}V
(H-(E+i\varepsilon))^{-1}\lambda^{1/2}V\big)\varphi_{k'}\rangle_\mathcal{H}\,,
\nonumber\\[1ex]
&&E=\omega(k)=\omega(k')\,,
\end{eqnarray}
with $\langle\cdot,\cdot\rangle_\mathcal{H}$ denoting the scalar
product of $\mathcal{H}$. The scattering rate $\sigma_\lambda(k,k')$ from
$k$ to $k'$ is the square of the $T$-matrix. Since in (\ref{2.6})
$k$ refers to the far future and $k'$ to the far past, one obtains
\begin{equation}\label{2.8}
\sigma_\lambda(k,k')=(2\pi)^3|T_\lambda(k',k)|^2
2\pi\delta(\omega(k)-\omega(k'))
\end{equation}
for $k\neq k'$. Recall that $\sigma_\lambda(k,k')$ is random through
the dependence on $V$.

In these notes we deal only with the average scattering rate,
$\langle\sigma_\lambda(k,k')\rangle_V$, for small $\lambda$. The
variance of $\sigma_\lambda$ is also of great interest and has been
studied both through experiments and theoretically \cite{BF94}.

To
be able to perform the Gaussian average we expand the resolvent of (\ref{2.4})
into the Born series as
\begin{equation}\label{2.8a}
T_\lambda(k',k)=\sum^{\infty}_{n=0}(-1)^n\langle
\varphi_{k'},\lambda^{1/2}V(G_{E+}\lambda^{1/2}V)^n\varphi_k\rangle_\mathcal{H}\,.
\end{equation}
Here $G_{E+}$ is the free Green's function,
\begin{equation}\label{2.8b}
G_{E+}=\lim_{\varepsilon\to 0}\big(H_0-(E+i\epsilon)\big)^{-1}\,,\quad
H_0=-\frac{1}{2}\Delta\,.
\end{equation}
Inserting (\ref{2.8a}) in (\ref{2.8}) one obtains
\begin{equation}\label{2.9}
\langle\sigma_\lambda(k,k')\rangle_V=
(2\pi)^3\langle\big|\sum^{\infty}_{n=0}(-1)^n\langle\varphi_{k'},
\lambda^{1/2}V(G_{E+}\lambda^{1/2}V)^n\varphi_k\rangle_\mathcal{H}\big|^2\rangle_V
2\pi\delta(\omega(k)-\omega(k'))\,.
\end{equation}
At this stage the average over $V$ can be carried out explicitly.
The result is most concisely expressed diagrammatically.
\begin{figure}[!ht]
    \begin{minipage}{0.5\textwidth}
    \centering
        \begin{psfrags}
                \psfrag{k}[][][1]{$k'$}
                \psfrag{ks}[][][1]{$k$}
                \psfrag{g1}[][][1]{$g_1$}
                \psfrag{k1}[][][1]{$k_1$}
                \psfrag{k2}[][][1]{$k_2$}
                \psfrag{k3}[][][1]{$k_3$}
                \includegraphics[height=2.25cm]{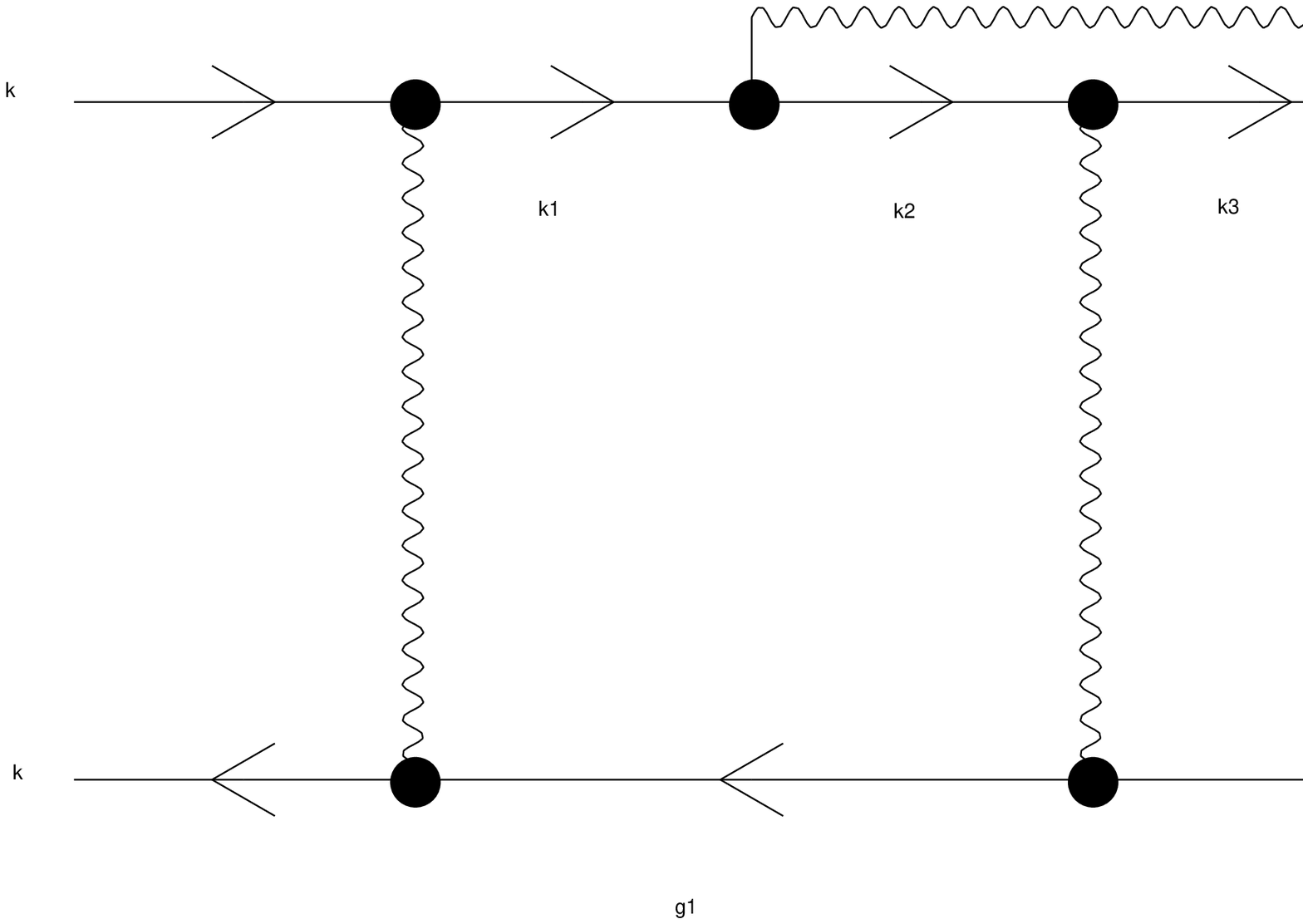}
            \end{psfrags}
    \end{minipage}
    \begin{minipage}{0.5\textwidth}
        \centering
        \raisebox{-21pt}{
            \begin{psfrags}
                \psfrag{k}[][][1]{$k_3$}
                \subfloat[]{\includegraphics[width=1.6cm]{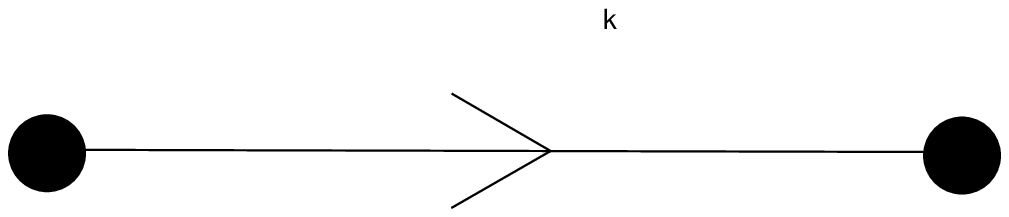}}
            \end{psfrags}
                \hspace{1cm}
            \begin{psfrags}
                \psfrag{k1}[][][1]{$k_1$}
                \psfrag{k2}[][][1]{$k_2$}
                \subfloat[]{\includegraphics[width=2cm]{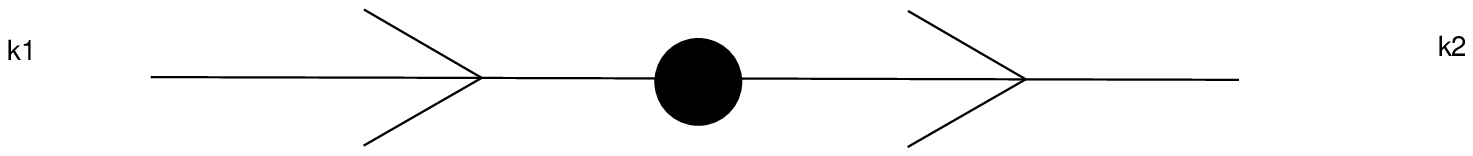}}
            \end{psfrags}
        }\\
        \vspace{18pt}\hspace{-11pt}
        \subfloat[]{\includegraphics[width=1.6cm]{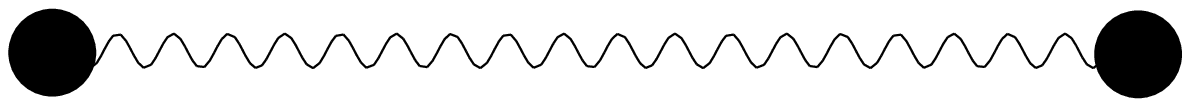}}
        \hspace{1cm}\hspace{5pt}
        \subfloat[]{\includegraphics[width=1.38cm]{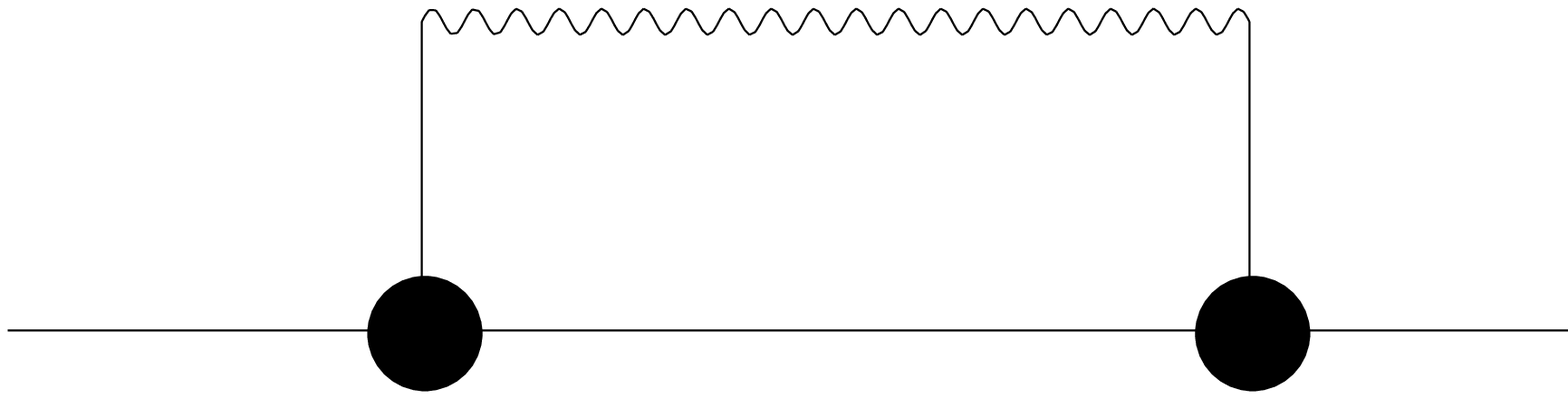}}
    \end{minipage}
\caption{A diagram and its basic building blocks.}
\end{figure}\\
We draw two horizontal lines, see Fig.~1, the top one is directed to
the right and the bottom one towards the left. They have external
momenta $k',k$. The top line carries $n_+$ vertices,
$n_+=1,2,\ldots$, and corresponding internal momenta
$k_1,\ldots,k_{n_+ -1}$. Each bond, see (a), with momentum $k$
carries the Green's function
$G_{E+}(k)=(\omega(k)-(E+i\varepsilon))^{-1}$. Each vertex, see (b),
with momenta $k_1,k_2$ carries the potential
$\lambda^{1/2}(2\pi)^{-3/2}\widehat{V}(k_1-k_2)$. Correspondingly,
the bottom line carries $n_{-}$ vertices, $n_-=1,2,\ldots$, and
internal momentum $g_1,\ldots,g_{n_--1}$. Each bond with momenta $g$
carries the Green's function $G_{E_-}(g)=G_{E_+}(g)^\ast$ and each
vertex with momenta $g_1,g_2$ carries the potential
$\lambda^{1/2}(2\pi)^{-3/2}\widehat{V}(g_1-g_2)$. The resulting
expression is integrated over all internal momenta
$k_1,\ldots,k_{n_+-1}$, $g_1,\ldots,g_{n_--1}$. The Gaussian average
generates a sum over all pairings, provided $n_++ n_-$ is even.
Otherwise the average vanishes. A pairing is indicated by a wavy
line, see (c).

For obvious reasons the building block (d) is called a gate. The sum
over all gates can be performed thereby modifying the free Green's
function (or propagator) to the effective medium Green's function as
\[
    \sum_{n=0}^{\infty} \Big( \raisebox{-3pt}{\includegraphics[width=1.6cm]{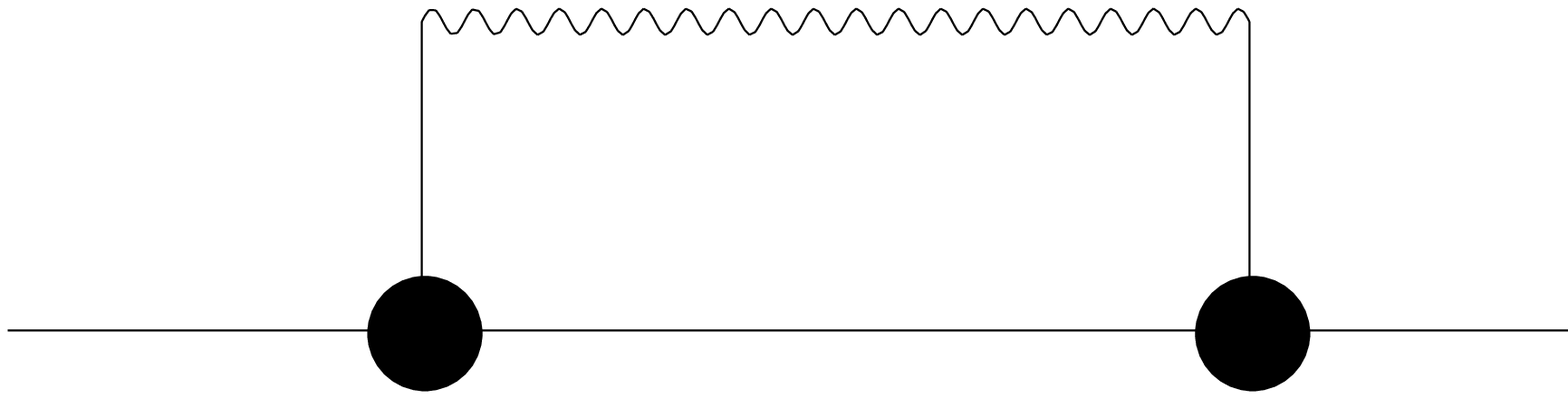}}\,\,\Big)^n =
    \raisebox{2pt}{\includegraphics[width=0.8cm]{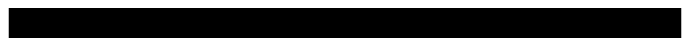}}\quad.
\]
Thus the diagrams remain as in Fig.~1 only the light lines are
replaced by thick lines. Because of the spatial cut-off in the
potential the effective medium Green's function is not as explicit as in
the translation invariant case.

In the following we study two particular classes of diagrams, namely
the ladder diagrams and the maximally crossed diagrams,
schematically represented in Fig. 2.
\begin{figure}[!ht]\centering
    \subfloat{
        \begin{psfrags}
            \psfrag{k}[][][1]{$k'$}
            \psfrag{ks}[][][1]{$k$}
                \includegraphics[height=1.837cm]{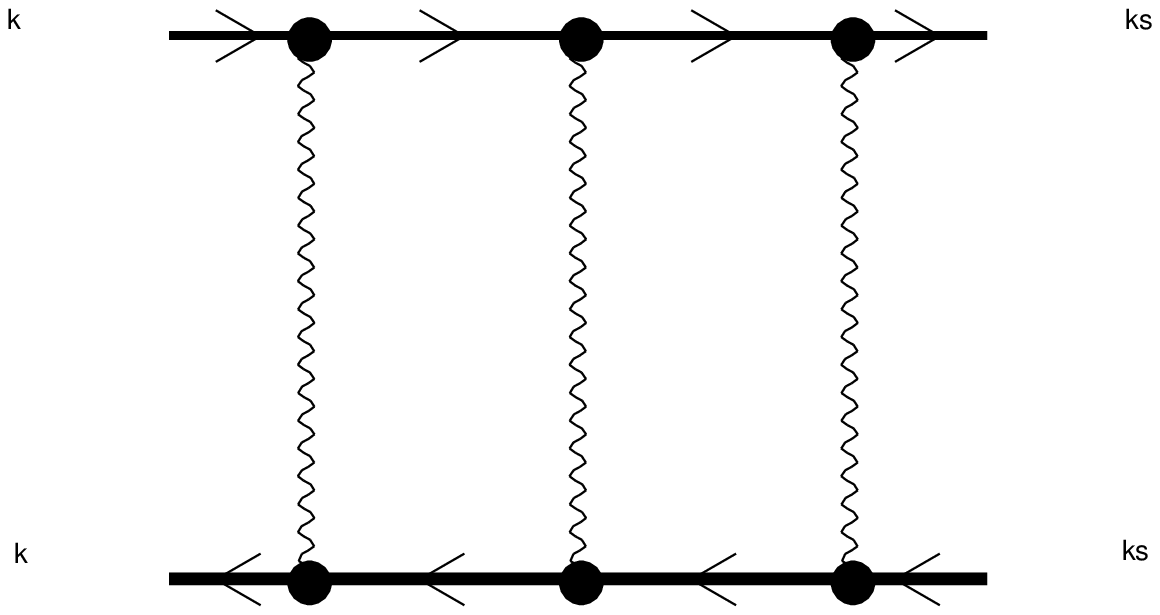}
        \end{psfrags}}
    \hspace{3cm}
    \subfloat{
        \begin{psfrags}
            \psfrag{k}[][][1]{$k'$}
            \psfrag{ks}[][][1]{$k$}
                \includegraphics[height=1.837cm]{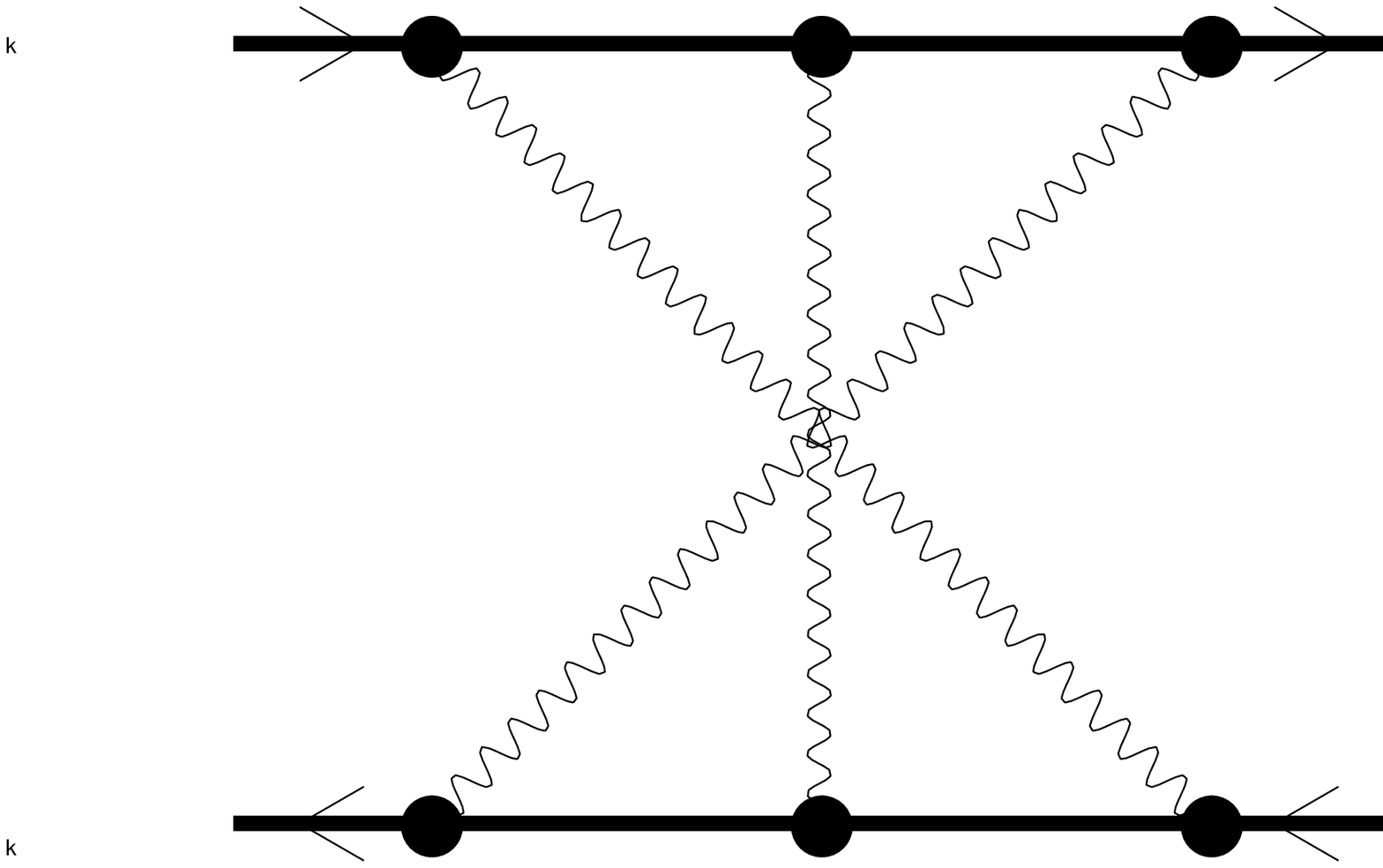}
        \end{psfrags}}
    \caption{A ladder diagram and a maximally crossed diagram with effective medium propagators.}
\end{figure}\\
We denote by $I_\mathrm{lad}(k',k;\lambda)$ the sum over all ladder
diagrams with external momenta $k'$ and $k$ and by
$I_\mathrm{max}(k',k;\lambda)$ the sum over all maximally crossed
 diagrams with external momenta $k'$ and $k$.

\section{Scattering rate from the Boltzmann equation}\label{sec.3}
\setcounter{equation}{0}

For a weak random potential one can approximate the true wave
dynamics by the stochastic motion of a fictitious classical
particle. It has position $q\in\mathbb{R}^3$, momentum
$p\in\mathbb{R}^3$, and kinetic energy $\omega(p)$. It moves with
constant momentum and changes its momentum at random times subject
to the constraint of constant energy. More precisely, given the
current location $q$ and momentum $p$, in the short time interval
$dt$ the probability for the momentum to be scattered into the
volume element $dp'$ is given by
\begin{equation}\label{3.1}
\rho(q)(2\pi)^{-3/2}\widehat{\theta}(p-p') 2\pi
\delta(\omega(p)-\omega(p'))
\end{equation}
independently for each small time interval. The corresponding
stochastic process is determined by the backward generator $L$,
which is defined through
\begin{eqnarray}\label{3.2}
&&\hspace{-20pt}Lf(q,p)=\nabla_p\,\omega(p)\cdot\nabla_q f(q,p)\\
&&\hspace{40pt}+\rho(q)\int
dp'2\pi\delta(\omega(p)-\omega(p'))(2\pi)^{-3/2}\widehat{\theta}(p-p')
\big(f(q,p')-f(q,p)\big)\nonumber
\end{eqnarray}
as a linear operator acting on functions
$f:\mathbb{R}^3\times\mathbb{R}^3\to\mathbb{C}$. The transition
probability from $(q,p)$ to $dq'dp'$ in time $t$, $t\geq 0$, is then
given by the integral kernel of $e^{Lt}$, denoted by
\begin{equation}\label{3.3}
e^{Lt}(q,p|dq'dp')\,,\quad t\geq 0\,.
\end{equation}
Clearly, outside of $\mathrm{supp}\rho$ the collision operator vanishes and
the particle moves freely.

The stationary scattering situation is easily modelled. We choose
$w\in\mathbb{R}^3$ away from $\mathrm{supp} \rho$ and a plane
$\Lambda_{w,\widehat{k}}$ through $w$ and orthogonal to the incoming
unit wave vector $\widehat{k}=k/|k|$ and not intersecting
$\mathrm{supp} \rho$. The initial position of the
particle is uniformly distributed on $\Lambda_{w,\widehat{k}}$ and
the initial velocity equals $\nabla\omega(k)$. The corresponding
initial measure $\mu_k$ imposes a uniform flux at momentum $k$
through the plane $\Lambda_{w,\widehat{k}}$, i.e.
\begin{equation}\label{3.4}
\mu_k(dqdp)=\delta(p-k)\delta((x-w)\cdot\widehat{k})
|\widehat{k}\cdot\nabla\omega(p)|dqdp\,.
\end{equation}
The Boltzmann scattering rate from $k$ to $k'$ is then given by
\begin{equation}\label{3.5}
2\pi\delta(\omega(k)-\omega(k'))\sigma_\mathrm{B}(k,k')=
\lim_{t\to\infty}\int\int\mu_k(dqdp)e^{Lt}(q,p|dq'dp')\delta(p'-k')
\end{equation}
for $k\neq k'$. For $k=k'$ there is a $\delta$-contribution coming
from those paths which pass the potential without any scattering. In
our set-up this $\delta$-function has infinite weight. More
meaningful would be to restrict $\mu_k(dqdp)$ to those initial
conditions for which the particle actually hits $\mathrm{supp}
\rho$. But the strict forward scattering $k'=k$ is of no concern to
us here, anyhow. The right side of (\ref{3.5}) is proportional to
$\delta(\omega(k)-\omega(k'))$. It is convenient to remove this
overall factor. On the energy shell, $\sigma_\mathrm{B}(k,k')$ is
uniquely defined and depends smoothly on its arguments.

In (\ref{3.5}) only the momentum of the scattered particle is
resolved. The limit $t\to\infty$ is needed, so that for any initial
condition on the plane $\Lambda_{w,\widehat{k}}$ the particle has
escaped from the scattering region.

We rewrite (\ref{3.5}) in a slightly more convenient form by
splitting the generator as
\begin{equation}\label{3.5a}
L=L_0+L_1
\end{equation}
with
\begin{equation}\label{3.6}
L_1 f(q,p)=\rho(q)\int dp'2\pi\delta(\omega(p)-\omega(p'))
(2\pi)^{-3/2}\widehat{\theta}(p-p')f(q,p')\,.
\end{equation}
Then
\begin{equation}\label{3.7}
e^{Lt}=e^{L_0t}+\int^t_0 ds e^{L_0s}L_1 e^{L_0(t-s)} +\int^t_0 ds
\int^s_0 ds' e^{L_0s'}L_1 e^{L(s-s')}L_1 e^{L_0(t-s)}\,.
\end{equation}
Let
\begin{equation}\label{3.8}
\nu(p)=\int dp'
2\pi\delta(\omega(p)-\omega(p'))(2\pi)^{-3/2}\widehat{\theta}(p-p')
\end{equation}
be the total cross section at momentum $p$ and set
\begin{eqnarray}\label{3.9}
&&\hspace{-36pt}f^+_k(q,p)=\delta(p-k)\exp\Big[-\nu(p)\int^\infty_0
dt \rho(q+\nabla\omega(p)t)\Big]\,,
\nonumber\\
&&\hspace{-36pt}f^-_k(q,p)=\delta(p-k)\exp\Big[-\nu(p)\int_{-\infty}^0
dt \rho(q+\nabla\omega(p)t)\Big]\,.
\end{eqnarray}

Inserting (\ref{3.7}) in (\ref{3.5}), the first term does not
contribute, since $k\neq k'$, and the second and third term have
limits, since $\mathrm{supp} \rho$ is bounded. Then
\begin{equation}\label{3.10}
2\pi\delta(\omega(k)-\omega(k'))\sigma_\mathrm{B}(k,k')= \langle
f^-_k,L_1 f^+_{k'}\rangle +\int^\infty_0 dt\langle f^-_k,L_1
e^{Lt}L_1 f^+_{k'}\rangle
\end{equation}
with $\langle\cdot,\cdot\rangle$ denoting the scalar product in
$L^2(\mathbb{R}^3\times\mathbb{R}^3,dqdp)$. The time integral
converges, since the probability to stay inside $\mathrm{supp} \rho$
decays exponentially.

\section{Summation over ladder diagrams}\label{sec.4}
\setcounter{equation}{0}

The goal of this section is to establish that for
$\omega(k)=\omega(k')$ it holds that
\begin{equation}\label{4.1}
\lim_{\lambda\to 0}\lambda^2(2\pi)^3
I_{\textrm{lad}}(k',k;\lambda)=\sigma_\mathrm{B}(k,k')\,.
\end{equation}
The prefactor $\lambda^2$ balances the cross section of the
scattering region which is of order $\lambda^{-2}$.

We set
\begin{equation}\label{4.2}
H_\pm(k_1)=\mp i\int dk_2(2\pi)^{-3/2}\widehat{\theta}(k_1-k_2)
G_{E\pm}(k_2)
\end{equation}
and note that
\begin{equation}\label{4.3}
H_+(k)+H_-(k)=\nu(k)\,.
\end{equation}

The gate
\raisebox{-10pt}{
\begin{psfrags}
    \psfrag{k1}[][][1]{$k_1$}
    \psfrag{k2}[][][1]{$k_2$}
    \psfrag{k3}[][][1]{$k_3$}
        \includegraphics[width=2cm]{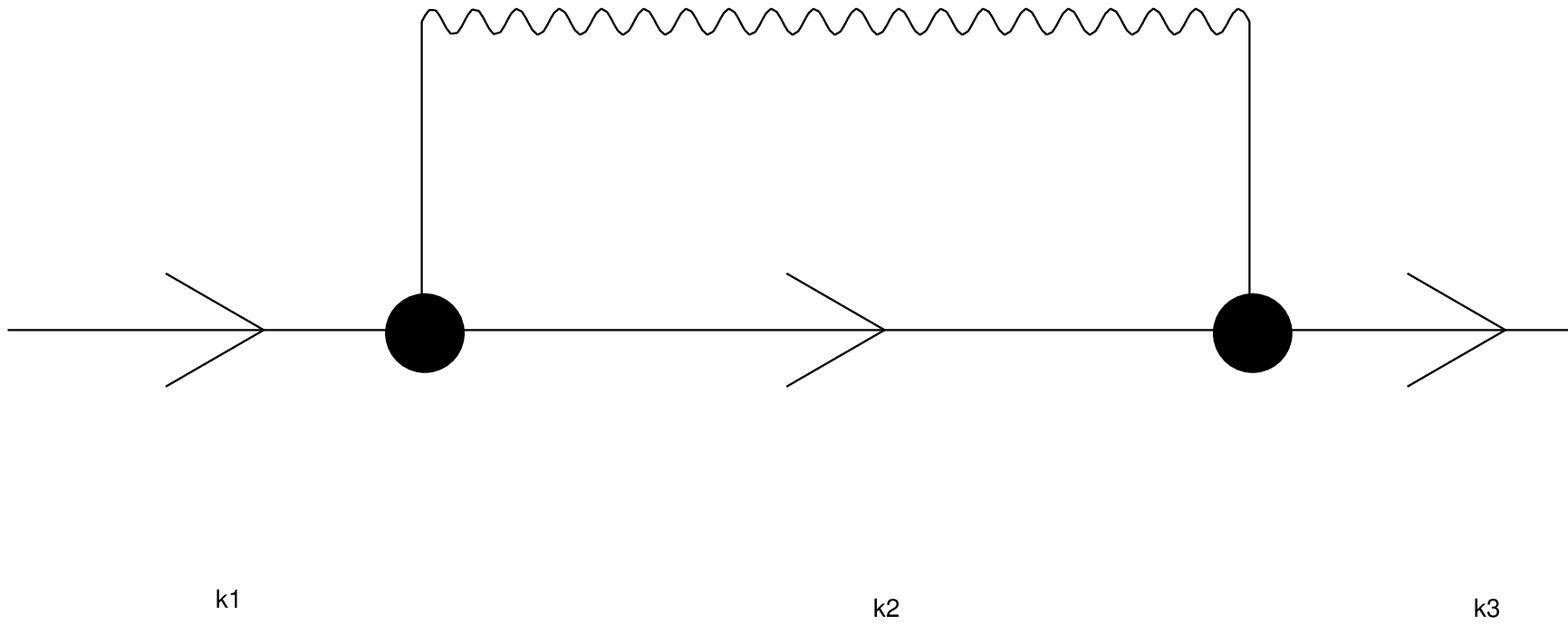}
\end{psfrags}}\,\,
corresponds to the integral operator
\begin{eqnarray}\label{4.4}
&&\hspace{-50pt}\int dk_2 G_{E+}(k_1)G_{E+}(k_2)\lambda(2\pi)^{-3}
\langle\widehat{V}(k_1-k_2)\widehat{V}(k_2-k_3)\rangle_V\nonumber\\
&&\hspace{-16pt}=\lambda(2\pi)^{-3}G_{E+}(k_1)\int
dk_2\widehat{\theta}(k_1-k_2)G_{E+}(k_2)\widehat{\rho}_\lambda(k_1-k_3)\nonumber\\
&&\hspace{-16pt}=i\lambda(2\pi)^{-3/2}G_{E+}(k_1)H_+(k_1)\widehat{\rho}_\lambda(k_1-k_3)\,.
\end{eqnarray}
Therefore the effective medium propagator,\,
\raisebox{+3pt}{
    \begin{psfrags}
        \psfrag{k1}[][][1]{$k_1$ }
        \psfrag{k}[][][1]{ $k$}
            \includegraphics[width=0.6cm]{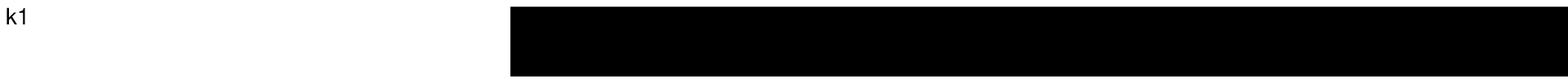}
    \end{psfrags}
}\,\,\,\,\,\,\,, is represented by the kernel
\begin{eqnarray}\label{4.5}
&&\hspace{-66pt}F_-(k_1,k;\lambda)=\delta(k_1-k)+
\sum^\infty_{m=2}\int dk_2\ldots dk_m\nonumber\\
&&\hspace{10pt}\prod^m_{j=1}\big\{G_{E+}(k_j) i
(2\pi)^{-3/2}H_+(k_j)\lambda\widehat{\rho}_\lambda(k_j-k_{j+1})\big\}\,,
\end{eqnarray}
where in (\ref{4.5}) we set $k_{m+1}=k$. We define
\begin{eqnarray}%XXX\label{4.5}
&&\hspace{-58pt}I_0(x;\lambda)=\int
dk_1\exp[i\lambda^{-1}(k_1-k)\cdot x]
F_-(k_1,k;\lambda)\nonumber\\
&&\hspace{-16pt}=\int dk_1 e^{i k_1\cdot x}\lambda^3 F_-(k+\lambda
k_1,k;\lambda)\,.
\end{eqnarray}

Shifting the other internal momenta by $k$ and rescaling by
$\lambda$ yields
\begin{eqnarray}\label{4.6}
&&\hspace{-58pt}I_0(x;\lambda)=1+\sum^\infty_{m=1}\int dk_1\ldots
dk_m e^{i k_1\cdot x}\prod^m_{j=1}\big\{\lambda G_{E+}(k+\lambda k_j) i(2\pi)^{-3/2}\nonumber\\
&&\hspace{26pt}H_+(k+\lambda
k_j)\widehat{\rho}(k_j-k_{j+1})\big\}\,,
\end{eqnarray}
where in (\ref{4.6}) we set $k_{m+1}=0$. $H_+$ is a smooth function. Thus $H_+(k+\lambda k_j)$ may be
replaced by $H_+(k)$. The Green's function $G_{E+}$ has the integral
representation
\begin{eqnarray}\label{4.7}
&&\hspace{-58pt}\lambda G_{E+}(k+\lambda k_1)=
i\lambda\int^\infty_0 dt e^{-it(\omega(k+\lambda k_1)-\omega(k)-i\varepsilon)}\nonumber\\
&&\hspace{24pt}\cong i\int^\infty_0 dt e^{-it(\nabla\omega(k)\cdot
k_1-i\varepsilon)}\,.
\end{eqnarray}
Inserting in (\ref{4.6}) yields
\begin{eqnarray}\label{4.8}
&&\hspace{-10pt}\lim_{\lambda\to 0}I_0(x;\lambda)\nonumber\\&&=
1+\sum^\infty_{m=1}\big(-H_+(k)\big)^m \int dk_1\ldots
dk_m(2\pi)^{-3m}
\int dx_1\ldots dx_m\rho(x_1)\ldots\rho(x_m)\nonumber\\
&&\hspace{10pt}\times\int^\infty_0 dt_1\ldots\int^\infty_0 dt_m
\exp\Big[-i\sum^m_{j=1}\big(t_j\nabla\omega(k)\cdot k_j
+(k_j-k_{j+1})\cdot x_j\big)+ik_1\cdot x\Big]\nonumber\\
&&\hspace{0pt}=1+\sum^\infty_{m=1}\big(-H_+(k)\big)^m
\int^\infty_0 dt_1\ldots\int^\infty_0 dt_m
\nonumber\\&&\hspace{60pt}\times\rho(x-\nabla\omega(k)t_1)\ldots
\rho(x-\nabla\omega(k)(t_1+\ldots +t_m))\nonumber\\
&&\hspace{0pt}=\exp\Big[-H_+(k)\int^0_{-\infty}dt
\rho(x+\nabla\omega(k)t)\Big]\,.
\end{eqnarray}

With this input we compute the single collision diagram with effective medium propagators,
\begin{figure}[!ht]
    \centering
        \begin{psfrags}
            \psfrag{k}[][][1]{$k'$}
            \psfrag{ks}[][][1]{$k$}
                \includegraphics[width=2.4cm]{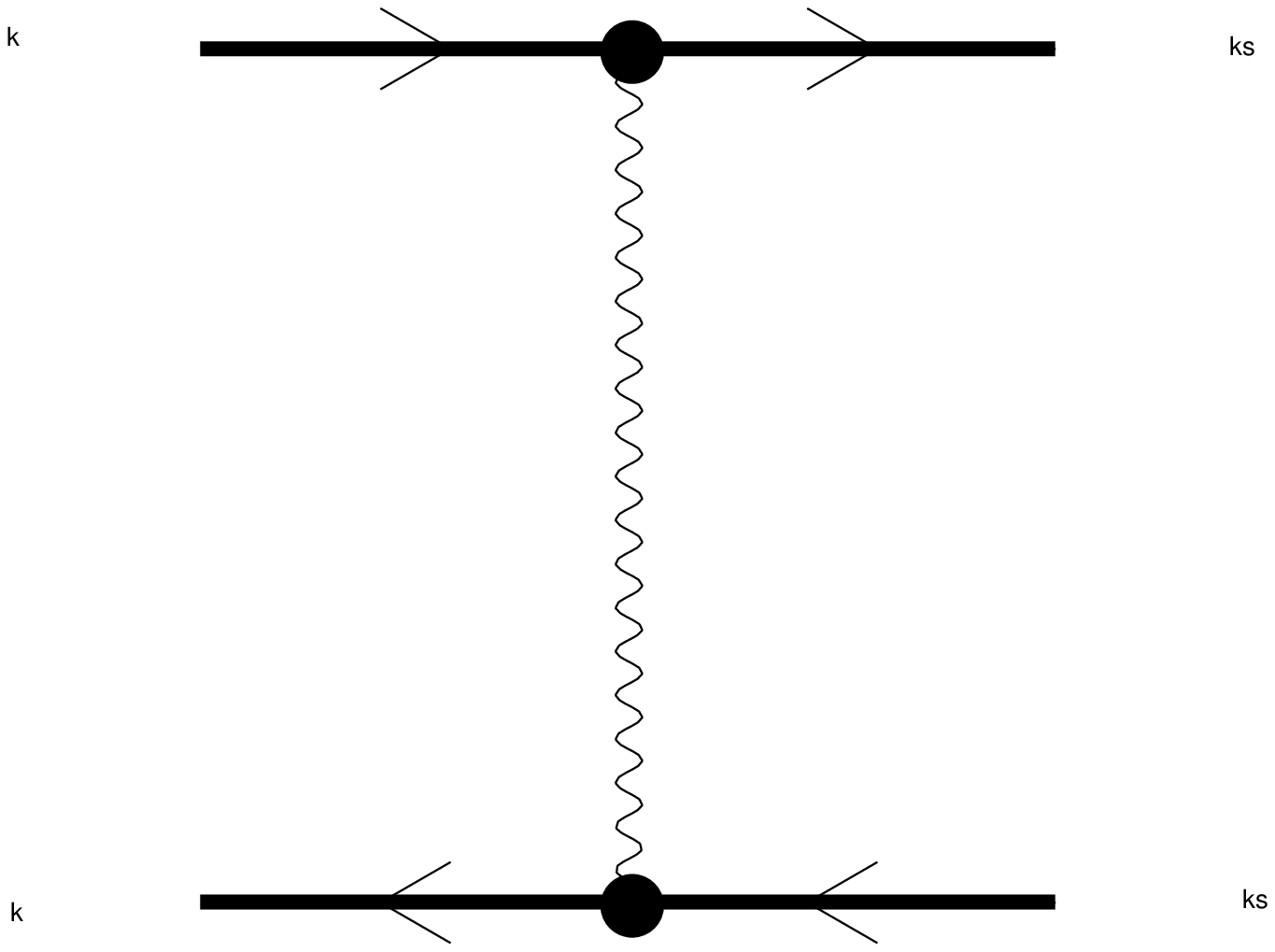}
        \end{psfrags}
\end{figure}\\
It is given by
\begin{eqnarray}\label{4.9}
&&\hspace{-38pt}I_1(\lambda)= \int dk_1 dk_2 dg_1
dg_2\lambda(2\pi)^{-3} \langle
\widehat{V}(k_1-k_2)\widehat{V}(g_1-g_2)^\ast\rangle_V\nonumber\\
&&\hspace{20pt} \times F_+(k_1,k'
;\lambda) F_-(k_2,k;\lambda)
F_+(g_1,k';\lambda)^\ast F_-(g_2,k;\lambda)^\ast\nonumber\\
&&\hspace{-8pt}=(2\pi)^{-3} \lambda\lambda^{-3} \int dk_1 dk_2 dg_1
dg_2 (2\pi)^{-3/2}\int dx\rho(x) \widehat{\theta}(k'-k+\lambda
k_1-\lambda k_2)\nonumber\\
&&\hspace{20pt}\times e^{-i k_1\cdot x}\lambda^3 F_-(k'+\lambda
k_1,k';\lambda)
\big(e^{-i g_1\cdot x}\lambda^3 F_-(k'+\lambda g_1,k';\lambda)\big)^\ast\nonumber\\
&&\hspace{20pt}\times e^{i k_2\cdot x}\lambda^3 F_+(k+\lambda
k_2,k;\lambda) \big(e^{i g_2\cdot x}\lambda^3 F_+(k+\lambda
g_2,k;\lambda)\big)^\ast\,.
\end{eqnarray}
Here $F_+$ is obtained from $F_-$ by substituting
$\widehat{\rho}_\lambda(-k_j+k_{j+1})$ for
$\widehat{\rho}_\lambda(k_j-k_{j+1})$ in (\ref{4.5}). Inserting
(\ref{4.8}) one obtains
\begin{eqnarray}\label{4.10}
&&\hspace{-30pt}I_1(\lambda)=(2\pi)^{-3}\lambda^{-2}\int dx
\rho(x)(2\pi)^{-3/2} \widehat{\theta}(k-k')\nonumber\\
&&\hspace{14pt} \exp\Big[-\big(H_+(k)+H_-(k)\big)
\int^0_{-\infty} dt \rho(x+\nabla\omega(k)t)\nonumber\\
&&\hspace{14pt}-\big(H_+(k')+H_-(k')\big)\int^\infty_0 dt
\rho(x+\nabla\omega(k')t)\Big]\big(1+\mathcal{O}(\lambda)\big)\,.
\end{eqnarray}
Using (\ref{4.3}) yields indeed
\begin{equation}\label{4.10a}
\lim_{\lambda\to 0}2\pi \delta(\omega(k)-\omega(k')) \lambda^2
(2\pi)^3 I_1(\lambda)=\langle f^-_k,L_1 f^+_{k'}\rangle\,.
\end{equation}

As next item we consider the ladder diagram with two collisions,
\begin{figure}[ht]
    \centering
        \begin{psfrags}
            \psfrag{k}[][][1]{$k'$}
            \psfrag{ks}[][][1]{$k$}
            \includegraphics[height=1.837cm]{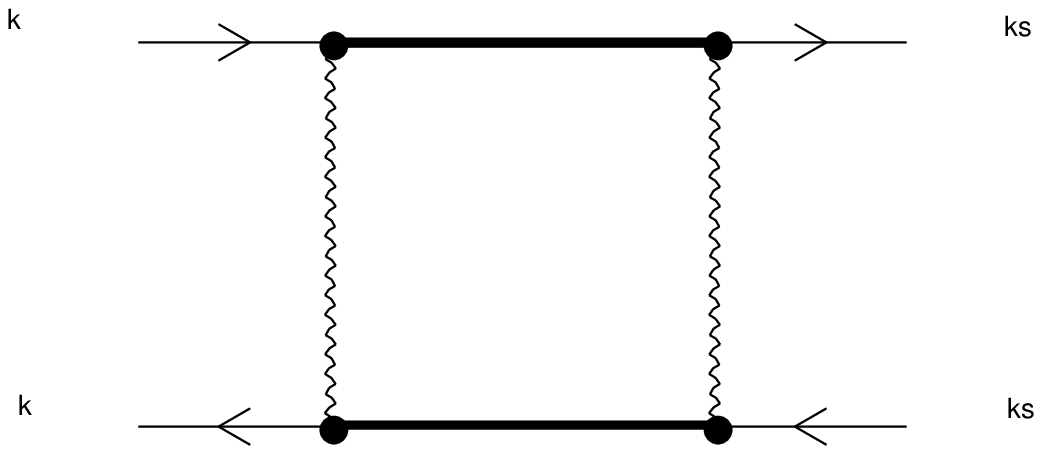}
        \end{psfrags}
\end{figure}\\
The external legs carry only the free propagator. Their gates can be summed to the effective medium propagator as in the first part of this section. The top line
has $m+2$ vertices and internal momenta $k_1,\ldots,k_{m+1}$,
$m=0,1,2,\ldots$. Correspondingly, the bottom line has $n+2$
vertices and internal momenta $g_1,\ldots,g_{n+1}$,
$n=0,1,2,\ldots$. The sum over all gates reads
\begin{eqnarray}\label{4.12}
&&\hspace{-20pt}I_2(\lambda)=\sum^\infty_{m=0}\sum^\infty_{n=0} \int
dk_1\ldots dk_{m+1}\int dg_1\ldots dg_{n+1}
\widehat{\theta}(k'-k_1)\widehat{\theta}(k_{m+1}-k)\\
&&\hspace{30pt}\times (2\pi)^{-6}\lambda\widehat{\rho}_\lambda(-k_1+g_1)
\lambda\widehat{\rho}_\lambda(-k_{m+1}+g_{n+1})^\ast\nonumber\\
&&\hspace{30pt}\times \prod^m_{j=1}\big\{\lambda(2\pi)^{-3/2}
G_{E+}(k_j)i H_+(k_j)\widehat{\rho}_\lambda(k_j-k_{j+1})\big\}\nonumber\\
&&\hspace{30pt}\times \prod^n_{j=1}\big\{\lambda(2\pi)^{-3/2}
G_{E-}(g_j)(-i)
H_-(g_j)\widehat{\rho}_\lambda(g_j-g_{j+1})^\ast\big\}
G_{E+}(k_{m+1})G_{E-}(g_{n+1})\,.\nonumber
\end{eqnarray}
We shift $k_j$ by $k_{m+1}$ and $g_j$ by $g_{n+1}$ and then
substitute $k_{m+1}=u+\frac{1}{2}v$, $g_{n+1}=u-\frac{1}{2}v$.
Finally $v$ is rescaled to $\lambda v$, $k_j$ to $\lambda k_j$, and
$g_j$ to $\lambda g_j$. Then (\ref{4.12}) becomes
\begin{eqnarray}\label{4.13}
&&\hspace{-20pt}I_2(\lambda)=\sum^\infty_{m=0}\sum^\infty_{n=0} \int
dk_1\ldots dk_m\int dg_1\ldots dg_n \int du dv
\widehat{\theta}(k'-\lambda k_1-u-\tfrac{1}{2}\lambda v)
\nonumber\\
&&\hspace{24pt}\times \widehat{\theta}(-k+u+\tfrac{1}{2}\lambda
v)(2\pi)^{-6}\lambda^2 \lambda^{-3}\widehat{\rho}(-k_1+g_1-v)
\widehat{\rho}(-v)^\ast\nonumber\\
&&\hspace{24pt}\times\prod^m_{j=1}\big\{\lambda(2\pi)^{-3/2}
G_{E+}(u+\tfrac{1}{2}\lambda v+\lambda k_j)i
H_+(u+\tfrac{1}{2}\lambda v+\lambda k_j)\widehat{\rho}(k_j-k_{j+1})\big\}\nonumber\\
&&\hspace{24pt}\times\prod^n_{j=1}\big\{\lambda(2\pi)^{-3/2}
G_{E-}(u-\tfrac{1}{2}\lambda v+\lambda g_j)(-i)
H_-(u-\tfrac{1}{2}\lambda v+\lambda g_j)\widehat{\rho}(g_j-g_{j+1})^\ast\big\}\nonumber\\
&&\hspace{24pt}\times G_{E+}(u+\tfrac{1}{2}\lambda
v)G_{E-}(u-\tfrac{1}{2}\lambda v)\,,
\end{eqnarray}
where in (\ref{4.13}) we set $k_{m+1}=0\,,\;g_{n+1}=0$.

We insert the approximation (\ref{4.7}) and switch to position
space,
\begin{eqnarray}\label{4.14}
&&\hspace{-8pt}I_2(\lambda)=\sum^\infty_{m=0}\sum^\infty_{n=0} \int
dk_1\ldots dk_m\int dg_1\ldots dg_n \int du dv
\widehat{\theta}(k'-u)
\widehat{\theta}(u-k)\nonumber\\
&&\hspace{24pt}\times\big(-H_+(u)\big)^m \big(-H_-(u)\big)^n \int^\infty_0
dt_1\ldots dt_m\int^\infty_0 ds_1\ldots ds_n \nonumber\\
&&\hspace{24pt}\times\exp\Big[-i \sum^m_{j=1}
t_j\big(\lambda^{-1}(\omega(u)-E-i\varepsilon)+\nabla\omega(u)\cdot
k_j+\tfrac{1}{2}\nabla\omega(u)\cdot v\big)\nonumber\\
&&\hspace{30pt}+i\sum^n_{j=1}s_j\big(\lambda^{-1}(\omega(u)-E+i\varepsilon)
+\nabla\omega(u)\cdot g_j- \tfrac{1}{2}\nabla\omega(u)\cdot
v\big)\Big]\nonumber\\
&&\hspace{24pt}\times\int dx_1\ldots
dx_m(2\pi)^{-3m}\rho(x_1)\ldots\rho(x_m) \int dy_1\ldots
dy_n(2\pi)^{-3n}\rho(y_1)\ldots\rho(y_n)\nonumber\\
&&\hspace{24pt}\times (2\pi)^{-6}\lambda^2 \lambda^{-3}(2\pi)^{-3}\int dx
dy \rho(x)\rho(y)\exp\Big[-i\sum^m_{j=1}x_j\cdot(k_j-k_{j+1})
\nonumber\\
&&\hspace{24pt}+i\sum^n_{j=1}y_j\cdot(g_j-g_{j+1})
-i x\cdot(-k_1+g_1 -v)-iy\cdot
v\Big]\nonumber\\
&&\hspace{24pt}\times G_{E+}(u+\tfrac{1}{2}\lambda v)
G_{E-}(u-\tfrac{1}{2}\lambda
v)\big(1+\mathcal{O}(\lambda)\big)\,,
\end{eqnarray}
where in (\ref{4.14}) we set $k_{m+1}=0,\,
g_{n+1}=0$.
We integrate over $k_1,\ldots,k_m$, $g_1,\ldots,g_n$ and substitute
\begin{equation}\label{4.15}
\sigma_j=\sum^j_{i=1} s_i\,,\quad \tau_j=\sum^j_{i=1} t_i\,.
\end{equation}
Then
\begin{eqnarray}\label{4.16}
&&\hspace{-8pt}I_2(\lambda)=\sum^\infty_{m=0}\sum^\infty_{n=0} \int
du dv (2\pi)^{-3} \widehat{\theta}(k'-u)
\widehat{\theta}(u-k)\big(-H_+(u)\big)^m \big(-H_-(u)\big)^n\nonumber\\
&&\hspace{24pt}\times\int_{0\leq\tau_1\ldots\leq\tau_m} d\tau_1\ldots
d\tau_m \int_{0\leq\sigma_1\ldots\leq\sigma_n} d\sigma_1\ldots
d\sigma_n (2\pi)^{-6}\lambda^2\lambda^{-3}\nonumber\\
&&\hspace{24pt}\times \int dx dy \rho(x)\rho(y)\prod^m_{j=1}\rho(x-\nabla\omega(u)\tau_j)
\prod^n_{j=1}\rho(y-\nabla\omega(u)\sigma_j)\nonumber\\
&&\hspace{24pt}\times \exp\big[i\big(E-\omega(u)\big)\lambda^{-1}(\tau_m-\sigma_n)
-i\tfrac{1}{2}\nabla\omega(u)\cdot v(\tau_m+\sigma_n)+i(x-y)\cdot
v\big]\nonumber\\[1ex]
&&\hspace{24pt}\times G_{E+}(u+\tfrac{1}{2}\lambda v)
G_{E-}(u-\tfrac{1}{2}\lambda v)\big(1+\mathcal{O}(\lambda)\big)\,.
\end{eqnarray}
We still have to take the limit $\lambda\to 0$ for the two remaining
Green's functions. Collecting all the $u,v$ dependence into the
smooth function $f$, this leads to
\begin{eqnarray}\label{4.17}
&&\hspace{-12pt}I_3(\lambda)=\lambda\int du dv f(u,v)
\exp\big[i\big(E-\omega(u)\big)\lambda^{-1}(\tau_m-\sigma_n)
-\tfrac{1}{2}i\nabla\omega(u)\cdot v(\tau_m+\sigma_n)\big]\nonumber\\
&&\hspace{40pt}\times G_{E+}(u+\tfrac{1}{2}\lambda v)
G_{E-}(u-\tfrac{1}{2}\lambda v)\nonumber\\
&&\hspace{18pt}=\lambda \int du dv f(u,v)\int^\infty_0 dt
\int^\infty_0 ds \nonumber\\
&&\hspace{34pt}
\times\exp\big[i\big(E-\omega(u)\big)\lambda^{-1}(\tau_m-\sigma_n)
-\tfrac{1}{2}i\nabla\omega(u)\cdot v(\tau_m+\sigma_n)\nonumber\\[1ex]
&&\hspace{58pt} \times i(t-s)\big(E-\omega(u)\big)-\varepsilon(t+s)
-\tfrac{1}{2}i\nabla\omega(u)\cdot
v\lambda(t+s)\big]\big(1+\mathcal{O}(\lambda)\big)\nonumber\\
&&\hspace{18pt}=\lambda\int du dv
f(u,v)\int^\infty_{\lambda^{-1}\tau_m} dt
\int^\infty_{\lambda^{-1}\sigma_n} ds
\exp\big[i\big(E-\omega(u)\big)(t-s)\nonumber\\[1ex]
&&\hspace{58pt}-\tfrac{1}{2}i\nabla\omega(u)\cdot
v(t+s)\big]\big(1+\mathcal{O}(\lambda)\big)\,.
\end{eqnarray}
We rotate $s,t$ by $\pi/4$ and rescale $t+s$ by $\lambda$. This
yields
\begin{equation}\label{4.18}
\lim_{\lambda\to 0} I_3(\lambda)=\int du dv f(u,v)
\int^\infty_{\max(\tau_m,\sigma_n)} dt e^{-i\nabla\omega(u)\cdot vt}
2\pi \delta(E-\omega(u))\,.
\end{equation}
Inserting (\ref{4.18}) in (\ref{4.16}) one extends the
$t$-integration from 0 to $\infty$, while $\tau_m\leq t$,
$\sigma_n\leq t$. Then
\begin{eqnarray}\label{4.19}
&&\hspace{-30pt}I_2(\lambda)=(2\pi)^{-3}\lambda^{-2}\int^\infty_0 dt
\int du dv (2\pi)^{-3} \widehat{\theta}(k'-u)\widehat{\theta}(u-k)
(2\pi)^{-3}\nonumber\\
&&\hspace{14pt}\times
\int dx dy \rho(x)\rho(y) 2\pi
\delta (E-\omega(u)) \exp\big[-i \nabla\omega(u)\cdot v t
+i(x-y)\cdot
v\big]\nonumber\\
&&\hspace{14pt}\times \exp\Big[-H_+(u)\int^t_0 ds
\rho(x-\nabla\omega(u)s)-H_-(u)\int^t_0 ds
\rho(y-\nabla\omega(u)s)\Big]\nonumber\\
&&\hspace{14pt} \times\big(1+\mathcal{O}(\lambda)\big)\,.
\end{eqnarray}
Let us set $f^0_k(q,p)=\delta(p-k)$. Then (\ref{4.19}) amounts to
\begin{equation}\label{4.20}
\lim_{\lambda\to 0} 2\pi \delta(\omega(k)-\omega(k')) \lambda^2
(2\pi)^3 I_2(\lambda)=\int^\infty_0 dt \langle f^0_k, L_1 e^{L_0
t}L_1 f^0_{k'}\rangle\,.
\end{equation}

Summing over the gates in the external legs amounts to replacing
$f^0_{k'}$ by $f^+_{k'}$ and $f^0_k$ by $f^-_k$. The ladder diagrams
with increasing number of rungs generate the time-dependent
perturbation theory for $e^{Lt}$. Thus we have verified the limit
(\ref{4.1}).

\section{Backscattering and maximally crossed diagrams}\label{sec.5}
\setcounter{equation}{0}

The first subleading contribution to $\langle
\sigma_\lambda(k,k')\rangle_V$ comes from the maximally crossed
diagrams. They are responsible for a narrow peak of width $\lambda$
centered at $k'=-k$ as a correction to $\sigma_\mathrm{B}(k,k')$.
Recall that $\varphi_k(q)=(2\pi)^{-3/2} e^{i k\cdot q}$. We define
\begin{equation}\label{5.1}
2\pi
\delta(\omega(k)-\omega(k'))\sigma_\mathrm{B}(k,k';\kappa)=(2\pi)^3
\int^\infty_0 dt \langle f^-_k\varphi_\kappa, L_1 e^{Lt}L_1
\varphi_\kappa f^+_{k'}\rangle
\end{equation}
and
\begin{equation}\label{5.2}
\sigma_{\textrm{back},k}(\kappa)=\sigma_\mathrm{B}(k,-k;\kappa)\,.
\end{equation}
Note that the first and last scattering event carries now an extra
phase factor. If one chooses $\omega(k)=\omega(-k+\lambda\kappa)$,
then the backscattering rate is given by
\begin{equation}\label{5.3}
\lim_{\lambda\to 0}(2\pi)^3 \lambda^2
I_\textrm{max}(k,-k+\lambda\kappa;\lambda)=
\sigma_{\textrm{back},k}(\kappa)\,.
\end{equation}
(\ref{5.3}) is our main result.

To verify our claim does not require a long computation. Let
$A_\textrm{max}$ be a particular maximally crossed diagram. If its
internal momenta at the lower line are $g_1,\ldots,g_m$, then we
substitute $-g_{m-j}$ for $g_m$. Then $A_\textrm{max}(k',k)=
\widetilde{A}_\textrm{lad}(k',k)$ where
$\widetilde{A}_\textrm{lad}(k',k)$ is the corresponding ladder
diagram with external momenta $k'$ and $k$ for the upper line and
$-k$ and $-k'$ for the lower line. In particular
$A_\textrm{max}(-k,k)=A_\textrm{lad}(-k,k)$, which implies
(\ref{1.3}). I.e. for precise backscattering at small $\lambda$ the
scattering rate is twice the one predicted by the Boltzmann
equation, upon omitting the single scattering contribution. This
result is independent of the shape function $\rho$.

For the fine structure we look $\lambda\kappa$ away from
backscattering and have to sum over all ladder diagrams, as
explained in Section 4, with the correspondingly modified external
momenta.

Let $k_1,k_2$, resp. $g_1,g_2$ be the internal momenta of the last
rung of the ladder diagram. Then the diagram beyond the last rung
equals
\begin{equation}\label{5.4}
I_r(k,-k+\lambda\kappa;\lambda)=\int dk_2
dg_2\lambda\langle\widehat{V}(k_1-k_2)\widehat{V}(g_1-g_2)^\ast\rangle_V
F_-(k_2,k-\lambda\kappa;\lambda) F_-(g_2,k;\lambda)^\ast\,.
\end{equation}

We shift $k_2$ to $k-\lambda\kappa+k_2$ and $g_2$ to $k+g_2$ and
rescale,
\begin{eqnarray}\label{5.5}
&&\hspace{-20pt}I_r(k,-k+\lambda\kappa;\lambda)= \lambda\lambda^{-3}
\int dk_2 dg_2 \widehat{\theta}(k_1-k-\lambda k_2+\lambda\kappa)
\widehat{\rho}(\lambda^{-1}(k_1-g_1)-k_2+g_2+\kappa)\nonumber\\
&&\hspace{24pt}\times \lambda^3 F_-(k-\lambda\kappa+\lambda
k_2,k-\lambda\kappa;\lambda)\lambda^3 F_-(k+\lambda
g_2,k;\lambda)^\ast \,.
\end{eqnarray}
Thus, compared to the ladder diagram, in position space the diagram
picks up the phase factor $e^{-i\kappa\cdot x}$. Repeating the
argument for the leftmost rung one concludes the validity of
(\ref{5.3}).

\section{Backscattering in the diffusive approximation}\label{sec.6}
\setcounter{equation}{0}

Let us consider the conventional slab geometry for which $\rho(r)=1$
if $r_3>0$ and $\rho(r)=0$ otherwise, $r=(r_1,r_2,r_3)$, and, for
the sake of illustration, choose $k=(0,0,k_3)$, $k_3>0$. The
scattering rate is proportional to the cross sectional area, through
which we divide by imposing that the first scattering is at
$r=(0,0,r_3)$ with $r_3>0$. Even with these simplifications the
inverse of $L$, as needed in (\ref{5.1}), is not so easily computed.
Therefore we approximate the motion between the first and last
scattering through a Brownian motion, i.e. $L$ through
$D(\omega)\Delta_r$, where $D(\omega)=D(\omega(k))$ is the diffusion
coefficient obtained from the transport equation (\ref{1.1}).
According to this equation the particle simply leaves the scattering
region upon hitting its boundary. Thus $\Delta_r$ is taken with
Dirichlet boundary conditions at $\{r_3=0\}$. Setting
$G_\mathrm{D}=(\Delta_r)^{-1}$ one has
\begin{equation}\label{6.1}
G_\mathrm{D}(r;r')=(4\pi|r-r'|)^{-1}-(4\pi|r-\widetilde{r}\,'|)^{-1}\,,\quad
r_3,r_3'>0\,,
\end{equation}
where $\widetilde{r}=(r_1,r_2,-r_3)$. The first and last scattering
are approximately a mean free path away from the boundary
$\{r_3=0\}$. Therefore the diffusion approximation is taken between
the points $r=(0,0,\ell^\ast)$ and $r'=(r_{\|},\ell^\ast)$ with the
mean free path $\ell^\ast=|\nabla_k\omega(k)|/\nu(k)$.

The backscattering in the diffusive approximation is thus given by
\begin{equation}\label{6.2}
\sigma_{\textrm{back},D}(k+\lambda\kappa)=\big(\nu(\omega)^2/D(\omega)\big)
\int_{\mathbb{R}^2}
dr_{\|}G_\mathrm{D}(0,0,\ell^\ast;r_{\|},\ell^\ast)
\big(1+\cos(\kappa\cdot r_{\|})\big)
\end{equation}
with $\nu(\omega)=\nu(\omega(k))$ and
$\kappa=(\kappa_1,\kappa_2,0)$. In the round bracket to the right
the ``1'' accounts for the incoherent scattering while the
``$\cos$'' expresses the coherent backscattering. Working out the
integral one obtains
\begin{equation}\label{6.3}
\sigma_{\textrm{back},D}(k+\lambda\kappa)=\big(\nu(\omega)^2/D(\omega)\big)
\ell^\ast\big(1+(2\ell^\ast|\kappa|)^{-1}
(1-\exp[-2\ell^\ast|\kappa|])\big)\,.
\end{equation}
We refer to \cite{AWM86} for a confirmation of (\ref{6.3}) in a
light scattering experiment.

\section{Light scattering}\label{sec.7}
\setcounter{equation}{0}

As an illustration we consider the wave equation with a random index
of refraction. Other wave equations can be handled in a similar
fashion. The wave field $\phi:\mathbb{R}^3\to\mathbb{R}$ is governed
by
\begin{equation}\label{7.1}
\frac{\partial^2}{\partial t^2}\phi=c(x)^2\Delta\phi
\end{equation}
and we set
\begin{equation}\label{7.2}
c(x)=1+\sqrt{\lambda}V(x)
\end{equation}
with $V$ from (\ref{2.3}). The Gaussian statistics looks unphysical
because $c(x)$ will have regions where it is negative. However for
a, say, bounded random potential only the Gaussian part persists for
small $\lambda$ \cite{ESY05,JS05}.

The general strategy is to rewrite (\ref{7.1}) in the form of a
Schr\"{o}dinger equation and then to apply the results from before.
Such a procedure is not unique and we adopt the one employed in
\cite{JS05}, see also \cite{S05}.

The dispersion relation for (\ref{7.1}) at $\lambda=0$ is
\begin{equation}\label{7.3}
\omega(k)=|k|\,.
\end{equation}
Again, more general dispersions can be handled by our method. If we
consider $\omega$ as multiplication in Fourier space, then the
corresponding operator in position space is denoted by $\Omega$. We
introduce the two-component ``wave function''
\begin{equation}\label{7.4}
\psi=(\psi^+,\psi^-)\,,
\end{equation}
where
\begin{equation}\label{7.5}
\psi^\pm(x)=\frac{1}{\sqrt{2}}\Big(\Omega\phi(x)\pm
i\big(1+\sqrt{\lambda}V(x)\big)^{-1}\dot{\phi}(x)\Big)\,.
\end{equation}
Then (\ref{7.1}) is rewritten as
\begin{equation}\label{7.6}
i \frac{\partial}{\partial t}
\begin{pmatrix}
\psi_+ \\\psi_- \\
\end{pmatrix}=
\begin{pmatrix}
\Omega & 0 \\0 & -\Omega \\
\end{pmatrix}
\begin{pmatrix}
\psi_+ \\\psi_- \\
\end{pmatrix} +\sqrt{\lambda}
\begin{pmatrix}
  V\Omega+\Omega V & V\Omega-\Omega V \\
  -V\Omega+\Omega V & -V\Omega-\Omega V \\
\end{pmatrix}
\begin{pmatrix}
\psi_+ \\\psi_- \\
\end{pmatrix}\,.
\end{equation}
Note that $\|\psi^+\|^2=\|\psi^-\|^2$ is the energy of the wave
field, hence conserved in time. We regard (\ref{7.6}) as an
evolution equation in $L^2(\mathbb{R}^3)\oplus L^2(\mathbb{R}^3)$.
The physical solutions are then the subspace defined by (\ref{7.5}).
The right-hand side of (\ref{7.6}) defines a self-adjoint operator,
thus the solution conserves the norm.

The only still missing input are the plane wave solution in case
$\lambda=0$. They are of the form $\phi_k(x)=(2\pi)^{-3/2}
\omega(k)^{-1}\exp[i(x\cdot k-\omega(k)t)]$, where the prefactor is
chosen such that the $\phi_k$'s are orthogonal in the energy norm.
Thus
\begin{equation}\label{7.7}
\psi^+(x)=(2\pi)^{-3/2} e^{ik\cdot x}\,,\quad \psi^-(x)=0\,.
\end{equation}

We have
\begin{eqnarray}\label{7.8}
&&\hspace{-20pt}(V\Omega\psi)\widehat{}(k)=(2\pi)^{-3/2}\int dk_1
\widehat{V}(k-k_1)\omega(k_1)\widehat{\psi}(k_1)\,,\nonumber\\
&&\hspace{-20pt}(\Omega V\psi)\widehat{
}(k)=(2\pi)^{-3/2}\omega(k)\int dk_1
\widehat{V}(k-k_1)\widehat{\psi}(k_1)\,.
\end{eqnarray}
In the perturbative expansion, compared to the Schr\"{o}dinger case,
a vertex carries an extra factor of $\omega$. This amounts to an
extra factor of $\omega^2$ in the collision rate. Thus the transport
equation for wave propagation has the backward generator
\begin{eqnarray}\label{7.9}
&&\hspace{-20pt}L_\mathrm{w} f(q,p)=\nabla\omega(p)\cdot\nabla_q
f(q,p)+\rho(q)\omega(p)^2 \int dp' 2\pi
\delta(\omega(p)-\omega(p'))\nonumber\\
&&\hspace{60pt}\times(2\pi)^{-3/2}\widehat{\theta}(p-p')\big(f(q,p')-f(q,p)\big)\,,
\end{eqnarray}
compare with (\ref{3.2}).

For the coherent backscattering of light one simply replaces in
(\ref{5.1}) $L$ by $L_\mathrm{w}$. In other words the collision rate
$\nu(p)$ is replaced by $\omega(p)^2\nu(p)$ and the propagation is
$\nabla\omega(k)=k/|k|$ rather than $k$. In the diffusion
approximation such fine details are no longer visible. Provided
$\ell^\ast$ is taken as the mean free path for light, the shape of
the coherent backscattering peak remains unaltered.

\end{document}